\documentclass[conference]{llncs}

\usepackage{cite}
\usepackage{amsmath,amssymb,amsfonts}
\usepackage{mathtools}
\usepackage{graphicx}
\usepackage{textcomp}
\usepackage{comment}
\usepackage{multirow}
\usepackage[table]{xcolor}
\usepackage[demo]{rotating}
\usepackage{tikz}
\usepackage{wrapfig}

\usepackage{hyperref}
\usepackage{url}

\usepackage{graphicx}
\usepackage{subfigure}

\usepackage{listings}
\usepackage{listing}
\usepackage{xspace}
\usetikzlibrary{arrows,automata}
\usepackage{subfigure}
\usepackage[firstpage]{draftwatermark}

\definecolor{dkblue}{rgb}{0,0.0,1}
\definecolor{prismgreen}{rgb}{0, 0.6, 0} 
\lstdefinelanguage{pseudo}{ 
        basicstyle=\color{black}\small\ttfamily, 
        keywords= 
        {global, var, IF, ELSE,FOR,EACH,false,WHILE, true, function,ENDIF,THEN}, 
        keywordstyle={\small\bfseries\color{dkblue}}, 
        numberstyle=\small\ttfamily\bfseries\color{black}, 
        comment=[l] {//}, morecomment=[s]{/*}{*/}, 
        commentstyle= \color{prismgreen}, 
        tabsize=4, 
        captionpos=b, 
        xleftmargin=20pt,
        xrightmargin=5pt,
        frame=single, 
        escapechar=@ 
}

\def\BibTeX{{\rm B\kern-.05em{\sc i\kern-.025em b}\kern-.08em
    T\kern-.1667em\lower.7ex\hbox{E}\kern-.125emX}}
\newcommand{\Clauses}{\mathcal{F}}
\newcommand{\SClauses}{\Clauses_S}
\newcommand{\HClauses}{\Clauses_H}
\newcommand{\bv}{\mathmbox{\textit{bv}}}
\newcommand{\ov}{\mathmbox{\textit{ov}}}
\newcommand{\rv}{\mathmbox{\textit{rv}}}
\newcommand{\uv}{\mathmbox{\textit{uv}}}
\newcommand{\cv}{\mathmbox{\textit{cv}}}
\newcommand{\idx}{\ensuremath{i}}

\newcommand{\arr}[1]{\stackrel{#1}{\longrightarrow}}
\newcommand{\TarTar}{\textsc{TarTar}\xspace}

\newcommand{\bvvar}{\mathit{v}}

\newboolean{arXiv} 
\setboolean{arXiv}{true} 

\SetWatermarkText{\hspace*{5in}\raisebox{5.5in}{\includegraphics[scale=0.1]{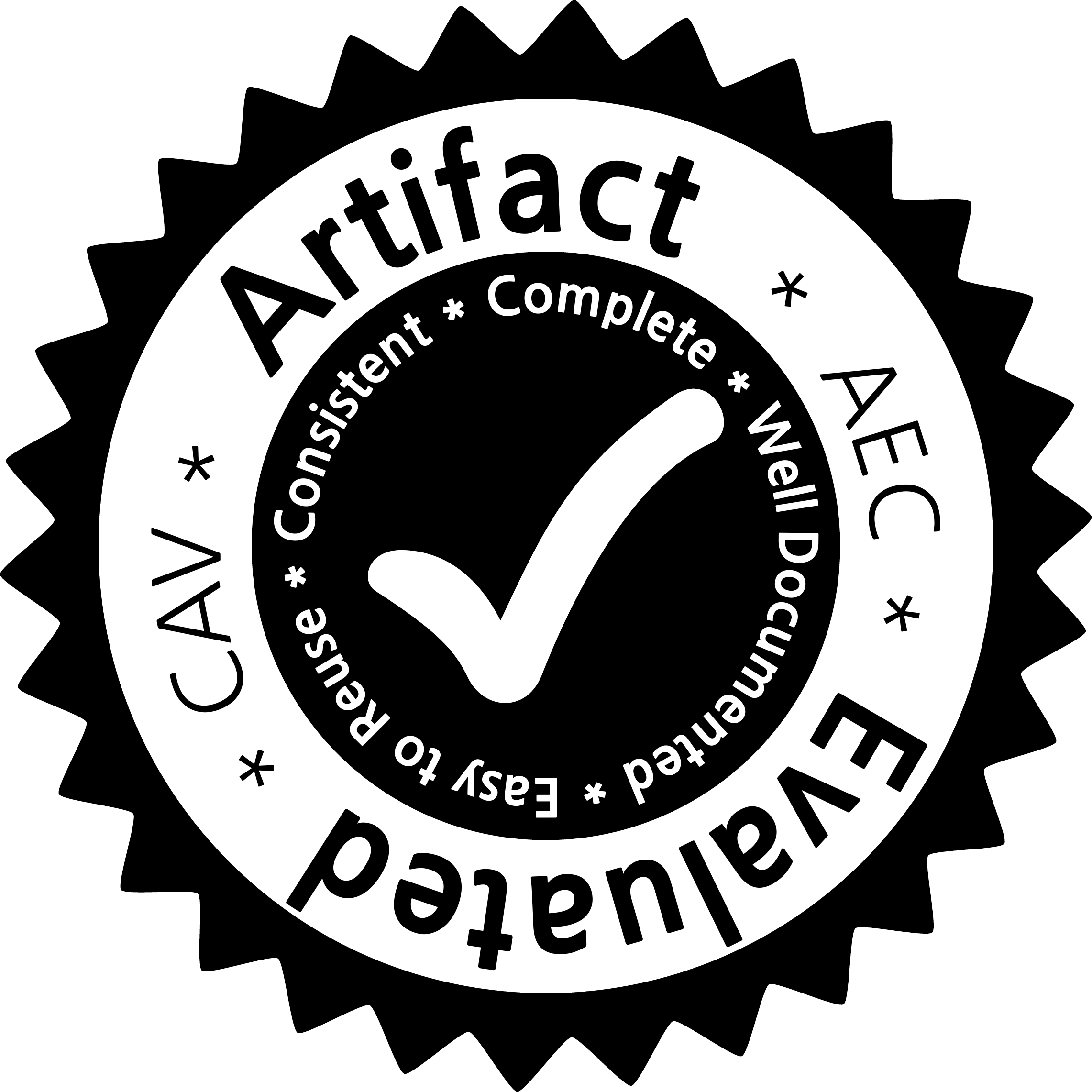}}}
\SetWatermarkAngle{0}

\begin{document}

\title{\TarTar: A Timed Automata Repair Tool}
\author{Martin~K\"olbl$^1$  \and Stefan~Leue$^1$ \and Thomas~Wies$^2$ }
\institute{\emph{$^1$University of Konstanz, Germany}, \emph{$^2$New York University, USA}}

\maketitle

\setcounter{secnumdepth}{2}
\setcounter{tocdepth}{2}

\begin{abstract}
We present \TarTar, an automatic repair analysis tool that, given a timed diagnostic trace (TDT) obtained during the model checking of a timed automaton model, suggests possible syntactic repairs of the analyzed model. The suggested repairs include modified values for clock bounds in location invariants and transition guards, adding or removing clock resets, etc. The proposed repairs are guaranteed to eliminate executability of the given TDT, while preserving the overall functional behavior of the system. We give insights into the design and architecture of \TarTar, and show that it can successfully repair 69\% of the seeded errors in system models taken from a diverse suite of case studies.
\end{abstract}

 \section{Introduction}
A reactive system with requirements pertaining to its timing behavior is often modeled as a network of timed automata (NTA)~\cite{BengtssonY03}.
Whether a timing requirement holds in an NTA can be analyzed by timed model checkers such as Uppaal~\cite{BenLarLaretal95} or opaal~\cite{DalHanetal11}. In case of a requirement violation, a model checker returns a timed counterexample, also called a timed diagnostic trace (TDT). 
Until now, developers must manually identify and correct such violations by analyzing the generated TDTs. It is therefore desirable to support this process by an automated tool
set that not only determines whether timing requirements are met, but also proposes syntactic repairs of the NTA in case they are not.

In~\cite{KoeLeuWie19} we presented an automated repair analysis that analyzes a TDT obtained from the violation 
of a timed safety property and returns syntactic repair suggestions that avoid the concrete executions of the TDT violating 
the property.
The analysis performs an additional admissibility check ensuring that the repaired model is functionally
equivalent with the original NTA, which means that no action traces are added or omitted by the repair.

 \begin{figure}[thb]
\hfill
\subfigure[Timed Automata \texttt{client}]
{\hspace{0.4cm}\includegraphics[width=3.7cm]{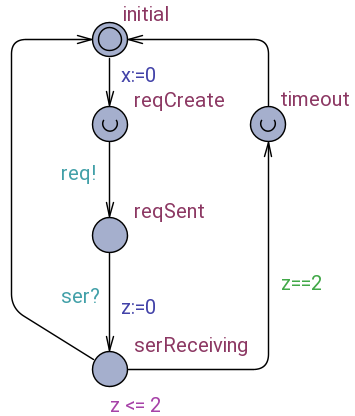}\label{fig:client}\hspace{0.4cm}}
\hfill
\subfigure[Timed Automata \texttt{db}]
{\includegraphics[width=3.2cm]{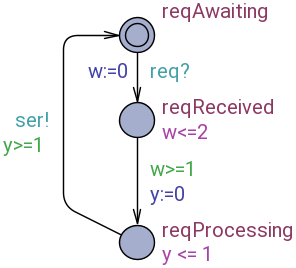}\label{fig:db}\hspace{0.2cm}}
\hfill
\subfigure[TDT \texttt{tdt}]
{\includegraphics[width=4.0cm]{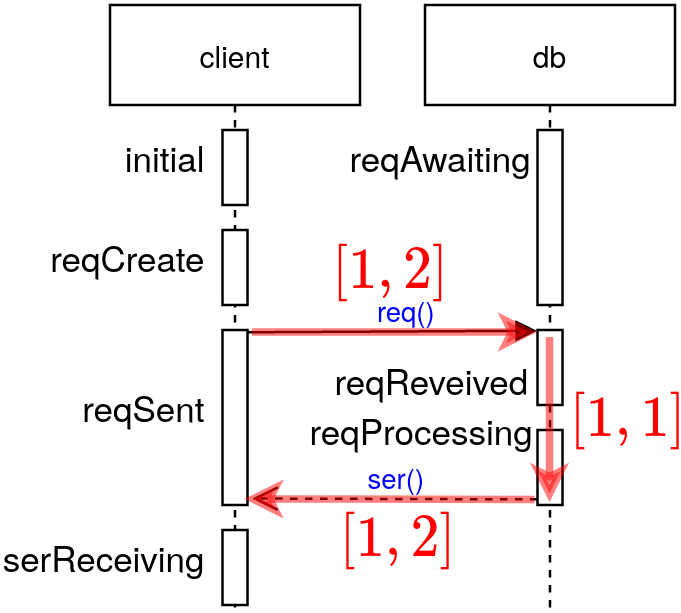}\label{fig:tdt}}
\hfill
\caption{Network of Timed Automata - Running Example}
\label{fig:running}
\end{figure}
To illustrate the repair analysis consider the NTA in Figures~\ref{fig:client} and~\ref{fig:db}. It describes a {\it client} that sends a request {\it req} to a database {\it db} and expects to receive a response {\it ser} within $4$ time units after sending the request. The client contains a clock $x$ that measures the time delay between the request creation and the receiving of a response in location {\it serReceiving}. The NTA allows to execute a TDT that violates the property, illustrated as a sequence diagram with time intervals in Figure~\ref{fig:tdt}. A time interval in the sequence diagram denotes the minimal and maximal time delay for the message transmission
and processing times in {\it db}, respectively.
The repair computation analyzes the TDT and produces several syntactic repairs to the NTA that avoid the property violation. In~\cite{KoeLeuWie19}, the computed repairs aim at the modification of clock bounds in location invariants and transition guards.
An example of such a repair is to reduce the bound in the time constraint $w \leq 2$ from $2$ to $1$. The modified bound constrains the maximal transmit time of the {\it req} message so that the resulting NTA receives all responses within the expected time. This repair eliminates the problematic executions of the TDT in the original NTA without changing the functional behavior of the system, which is confirmed by an admissibility test defined in~\cite{KoeLeuWie19}. However, in general, it may not be possible to repair the model using only clock bound alterations.

\paragraph{Contributions.} We present \TarTar~\cite{tartar}, which extends the initial prototype implementation of the clock bound repair analysis presented in~\cite{KoeLeuWie19} to a more comprehensive NTA repair tool. Specifically, the extended tool implements new analyses that can suggest a whole range of repairs in addition to clock bound variation, such as modifying comparison operators in constraints, clock references, clock resets, and location urgency. Examples of new repairs computed for the model in Figure~\ref{fig:running} are:
\begin{itemize}
    \item Exchanging the comparison operator in the constraint $w \geq 1$ to $w<1$ ensures that the time to send a request is below $1$ time unit.
    \item An exchange of clock $z$ in $z \leq 2$ with clock $y$ restricts the time of processing and receiving the response to at most 2 time units.
    \item To reset the clock $y$ on the previous transition instead ensures that the time for sending and processing the request is below $1$ time unit.
    \item Making the location {\it serReceiving} urgent reduces the time to receive a response to 0.
\end{itemize}
We call a repair admissible if the repaired system is functionally equivalent to the unrepaired system. 
The repair analysis implemented in \TarTar returns the complete set of admissible repairs.

The repair analysis combines concepts and algorithms from model checking, constraint solving, and automata theory. A real-time model checker is used to generate TDTs for a given NTA that violate a given timed safety property. \TarTar translates the TDT
into a linear real arithmetic constraint system. 
An SMT solver is used to compute a repair for the generated constraint system by solving a MaxSMT problem.  
An automata-based language equivalence test checks whether the repair is admissible in the NTA model. The collaboration between these subcomponents yields a complex tool architecture. We provide insights into the design and implementation of this architecture and the underlying infrastructure of supporting tools.
We evaluate the new repair analyses by applying \TarTar to a number of NTA models. We systematically inject different modifications in these correct models and compute repairs for the obtained faulty models, which results in at least one admissible repair for $69$\% of the TDTs. 

\paragraph{Related Work.}
Other tools exist that compute repairs.
The tool BugAssist~\cite{JosMaj11} analyzes C-code by solving a MaxSMT problem.
The tool ReAssert~\cite{DanGveetal11} checks a set of possible modification to repair broken unit tests.
Angelix~\cite{MecYiRoy16}, S3~\cite{BacHieLoetal17} and SemFix~\cite{NguQiRoyetal13} compute repairs by symbolic execution and constraint solving.
SketchFix~\cite{HuaZhaKhu18} is based on lazy candidate generation.
All tools are not repairing broken time constraints.
We are not aware of related work on tools for the repair of timed automata models. 
A more comprehensive overview of related work on automated repair is given in~\cite{GouPraRoy19}.
A discussion of work related to the foundations of our repair analysis can be found
in~\cite{KoeLeuWie19}.

\ifthenelse{\boolean{arXiv}}
{\section{Preliminary}
\subsection{Basic Repair Analysis}

We start with a brief summary of the repair analysis approach proposed in~\cite{KoeLeuWie19} and then discuss the extensions implemented in \TarTar.

\paragraph{Preliminaries.}
The timed automaton model that we use is adapted from~\cite{BengtssonY03}.
Given a set of \emph{clocks} $C$, we denote by
${\cal B}(C)$ the set of all \emph{clock constraints} over $C$, which are
conjunctions of \emph{atomic clock constraints} of the form $c \sim n$, where
$c \in C$, $\sim \in \{<,\leq, =, \geq, >\}$ and $n \in \mathbb{N}$.
A \emph{timed automaton (TA)} $T$ is a tuple $T = (L, l^0, C, \Sigma, \Theta, I)$ where
$L$ is a finite set of locations,
$l^0 \in L$ is an initial location,
$C$ is a finite set of clocks,
$\Sigma$ is a set of action labels,
$\Theta \subseteq_{\mathit{fin}} L \times {\cal B}(C) \times \Sigma \times 2^C
\times L$ is a set of \emph{actions}, and
$I: L \rightarrow {\cal B}(C)$ denotes a labeling of locations with clock constraints, 
referred to as location invariants.
For $\theta \in \Theta$ with $\theta = (l, g, a, r, l')$, we refer to
$g$ as the \emph{guard} of $\theta$ and to $r$ as its \emph{clock
  resets}.
An urgent location is a location that has to be left again without any delay in time~\cite{BouFahGuletal18}.
Urgent locations are syntactic sugar of Uppaal and can be expressed as an additional clock $p$ which is reset with entering the location and a location invariant $p=0$. A set $\textit{urgent}$ contains all urgent locations of $T$.

A repair analysis is possible for timed safety properties~\cite{BengtssonY03}.
A timed safety property $\Pi$ is encoded as a Boolean combination of atomic clock constraints and location predicates. A location predicate $@l$ with $l\in L$ holds in a state $(l', u)$ of a TA iff $l = l'$. 
A TA satisfies $\Pi$ if all reachable states satisfy $\Pi$. Otherwise, a timed model checker returns a TDT leading to a state that dissatisfies $\Pi$.

\paragraph{Repair Analysis}
The repair analysis analyzes the action sequence of the TDT that leads to a violation of property $\Pi$. The delay $\delta_i$ is the time between two actions $a_i$ and $a_{i+1}$ of the sequence to happen. We define the sequence as a symbolic timed trace. 

A \emph{symbolic timed trace} (STT) S of a TDT is a sequence of actions
$S = \theta_0,\dots,$ $\theta_{n-1}$. A \emph{realization} of $S$ is a
sequence of delay values $\delta_0,\dots,\delta_{n}$ such that there
exists states $s_0,\dots,s_n,s_{n+1}$ with $s_i \arr{\delta_{i}}
\arr{\theta_{i}} s_{i+1}$ for all $i \in [0,n)$ and $s_n
\arr{\delta_{n}} s_{n+1}$. We say that a STT
is \emph{feasible} if it has at least one realization.

Given an STT $S$, the repair analysis proceeds according to the following steps.

\paragraph{Step 1: Trace Encoding.} 
We first encode $S$ as a \emph{timed diagnostic trace constraint system} (TDTCS), expressed in linear real arithmetic. The satisfying assignments of the TDTCS precisely capture the feasible realizations of $S$. The TDTCS is given by the conjunction $\mathcal{T}$ of the following constraints:
\begin{align*}  
\mathmbox{\cal{C}}_0 \equiv \; &
\bigwedge\limits_{c \in C} c_{0} = 0 &
\text{(clock initialization)}\\
\mathmbox{\cal{A}} \equiv \; & 
\bigwedge\limits_{j \in [0,n]}
\delta_{j} \geq 0 & \text{(time advancement)}\\
\mathmbox{\cal{R}} \equiv \; &
\bigwedge\limits_{c \in \mathmbox{\textit{reset}}_j, }
c_{j+1} = 0 & \text{(clock resets)}\\
\mathmbox{\cal{U}} \equiv \; &
\bigwedge\limits_{l_j \in \mathmbox{\textit{urgent}} }
d_j = 0 & \text{(urgent location)}\\
\mathmbox{\cal{D}} \equiv \; &
\bigwedge\limits_{c \notin \mathmbox{\textit{reset}}_j} c_{j+1} = c_{j} + \delta_j & \text{(sojourn time)}\\
\mathmbox{\cal{I}} \equiv \; &
\bigwedge\limits_{(c \sim \beta) \in I(l_j)}
c \sim \beta
\wedge c + \delta_j \sim \beta
& \text{(location invariants)}\\
\mathmbox{\cal{G}} \equiv \; &
\bigwedge\limits_{(c \sim \beta) \in g_j}
c + \delta_j \sim \beta
& \text{(transition guards)}\\
\mathmbox{\cal{L}} \equiv \; & @l_n \land \bigwedge_{l
    \neq l_n} \neg @l & \text{(location predicates)}
\end{align*}

Let further $\mathmbox{\Phi} \equiv \Pi[\mathbf{c}_{n+1}/\mathbf{c}]$
where $\Pi[\mathbf{c}_{n+1}/\mathbf{c}]$ is obtained from property $\Pi$ by
substituting all occurrences of clocks $c \in C$ by $c_{n+1}$. Then the $\Pi$-extended TDTCS associated with $S$ is defined as
$\mathcal{T}^\Pi = \mathcal{T} \land \neg \Phi$.

\paragraph{Step 2: Repair Encoding.} 
    A system repair is a variation of clock constraints that no longer allows any execution of the TDT that leads to a property violation.
    We encode the clock variations by exchanging the clock constraints of $\mathmbox{\cal{T}}$ by modifiable constraints.
    Each constant $b_i$ used as a clock bound in a clock constraint is replaced by the expression $b_i + \bvvar_i$ where $\bvvar_i$ is a fresh variable indicating the change to $b_i$.
    For instance, a guard with the time constraint $c_j + \delta_j \sim \beta_i$ will be replaced by $c_j + \delta_j \sim \beta_i + \bvvar_i$. 
    For the clock bound analysis, we exchange all clock bounds in ${\cal I}$ and ${\cal G}$, resulting in a new constraint sets ${\cal I}^{\bv}$ and ${\cal G}^{\bv}$. 
    We then derive from $\mathcal{T}$ a constraint 
    system $\mathmbox{\cal{T}}^{bv}$ by replacing ${\cal I}$ and ${\cal G}$ with ${\cal I}^{\bv}$ and ${\cal G}^{\bv}$. The system $\mathmbox{\cal{T}}^{\bv}$ captures all realizations of the STT $S$ where the clock bounds of the underlying automaton have been modified according to the chosen values for the variables $\bvvar_i$.

\paragraph{Step 3: Repair Computation.}

    We use $\mathcal{T}^{bv}$ to derive an instance of the partial MaxSMT problem whose solutions yield candidate repairs for the timed automaton. The partial MaxSMT problem takes as input a finite set of assertion formulas belonging to a fixed first-order theory. The assertions are partitioned into hard $\HClauses$ and soft $\SClauses$ assertions.
    A solution to the problem is a maximal subset $\mathcal{F} \subseteq \SClauses$ such that $\HClauses \cup \mathcal{F}$ is satisfiable. 
    For our analysis, the hard constraints are given by
    \[\HClauses \equiv (\exists \delta_j, c_j.\, {\cal{T}}^{\bv}) \land (\forall \delta_j, c_j.\, \mathmbox{\cal{T}}^{\bv} \Rightarrow \Phi).\]
    This formula describes all assignments to the block bound variation variables $\bvvar_i$ such that the TDT no longer admits any realization that violates the property and such that at least one realization still exists.

    To obtain a repair with a minimal number of modified clock bounds, we utilize the ability of the partial MaxSMT problem to maximize the number of soft asserts that hold.
    To this end, let $\bvvar_1$ to be $\bvvar_n$ be an enumeration of all the clock bound variation variables, then we define the soft assertions as
\[
\SClauses \equiv 
\bvvar_1 = 0 \land \dots \land \bvvar_n = 0.
\]
Clearly, $\HClauses \land \SClauses$ with $\forall i.\bvvar^{\bv}_\idx = 0$  is not satisfiable because $\mathcal{T}^{bv} \land \SClauses$ is equisatisfiable with $\cal{T}$, and ${\cal{T}} \land \neg \Phi$ is satisfiable by assumption. However, if there exists at least one repair for $S$, then $\HClauses$ alone is satisfiable. In this case, the MaxSMT instance $\HClauses \cup \SClauses$ has at least one solution.

\paragraph{Step 4: Admissibility Check.}
A repair can affect the possible executions of an NTA. On the one hand, the repair removes the execution that violates the property, on the other hand, it can change the functional behavior of the system.
    We developed an admissibility check that ensures the functional equivalence of the original and repaired system in \cite{KoeLeuWie19}. The functionality of an NTA $T$ is captured by its untimed language $L_\mu$~\cite{AluDil94}, i.e., the sequences of actions observed in its runs.
    A repair is admissible iff the repaired NTA has the same untimed language as the original NTA. 
}
{}
\section{New Types of Repair Analyses}

The repair analysis presented in~\cite{KoeLeuWie19} and implemented in the prototype version of \TarTar 
encodes a TDT as a constraint system in linear real arithmetic. 
It computes syntactic correct modifications of the underlying NTA by introducing bound variation variables $\bvvar$.
For example, possible bound modifications for a clock bound $x \leq 2$ are expressed by a modified clock bound $x\leq 2+\bvvar$.
The repairs are computed by solving a partial SMT problem on the TDT constraint system, involving
soft-assert constraints on the bound variation variables.
No repair is computed whenever the soft assertion $\bvvar=0$ holds, otherwise the computed value of $\bvvar$
characterizes the repair. In the following we sketch the new types of repairs implemented in \TarTar. 
\ifthenelse{\boolean{arXiv}}
{}
{
For a more
comprehensive description, which space limitations do not allow us to provide here, we refer to~\cite{KoeLeuWie20}.
}

\paragraph{Operator Variation Repair Analysis.} 
This analysis is motivated by the assumption that a wrong comparison operator in a location
invariant or transition guard may cause a property violation.
We assume for the repair encoding that the operators $\sim$ are indexed according to 
their order in the sequence $\langle \; <, \leq, =, \geq, >\> \rangle$.
The possible repairs are encoded by a fresh variation variable $\bvvar^{ov}_\idx$ 
where the value of $\bvvar^{ov}_\idx$ is the index of the corresponding comparison operator.
If $x < 4$  is computed as a repair, then $\bvvar^{ov}_\idx = 1$.
\ifthenelse{\boolean{arXiv}}
{
We define appropriate operator variation constraints $\mathmbox{{\cal I}^{ov}}$ and $\mathmbox{{\cal G}^{\ov}}$ with the help of an n-ary exclusive or operation $\bigoplus\limits_{i = 0 \ldots n}f_i$ which is satisfied iff exactly one of the formulas $f_i$ is true:
\begin{align*}  
\mathmbox{{\cal I}^{\ov}} \equiv \; &
\bigwedge \limits_{(c \sim \beta) \in I(l_j)}
  \bigoplus\limits_{0 \leq k \leq 5}
  c\sim_k \beta \wedge
  c + \delta_j\sim_k \beta \wedge
   \bvvar^{\ov}_\idx = k.\\
\mathmbox{{\cal G}^{\ov}} \equiv \; &
\bigwedge \limits_{(c \sim \beta) \in g_j} 
\bigoplus\limits_{0 \leq k \leq 5} 
c + \delta_j\sim_k \beta \wedge \bvvar^{\ov}_\idx = k.
\end{align*}

}
{
}
Using this repair analysis, \TarTar finds two admissible repairs for the example in Figures~\ref{fig:client} 
and~\ref{fig:db} that replace the comparison operator in the clock constraint $w>=1$ by $<$ or $<=$, respectively.

\paragraph{Clock Reference Repair Analysis.}
This analysis aims to repair property violations resulting from errors that stem from the unintended use 
of a wrong clock variable.
We enumerate all the positions of clock variables in clock bound constraints using index $\idx$ and 
all clock variables using index $k$. We then introduce for every position $\idx$, a fresh variation 
variable $\bvvar^{\cv}_\idx$ whose value $k$ indicates the clock $c_k$ to be used at that position 
in the repaired model. 
For example, if $y \leq 2$ is a repaired constraint, where the position of $y$ in the constraint has index $3$ and clock $y$ has index $1$, then $\bvvar^{\cv}_3 = 1$.
\ifthenelse{\boolean{arXiv}}
{
We define the appropriate clock variation constraints $\mathmbox{{\cal I}^{\cv}}$ and $\mathmbox{{\cal G}^{\cv}}$:
\begin{align*}  
\mathmbox{{\cal I}^{\cv}} \equiv \; &
\bigwedge \limits_{(c \sim \beta) \in I(l_j)}
\bigoplus\limits_{0 \leq k \leq |C|} (c_{k}\sim
\beta)\wedge (c_{k} + \delta_j\sim \beta)\wedge(\bvvar^{\cv}_\idx = k)\\
\mathmbox{{\cal G}^{\cv}} \equiv \; &
\bigwedge \limits_{(c \sim \beta) \in g_j}
\bigoplus\limits_{0 \leq k \leq |C|} (c_{k} + \delta_j\sim
\beta)\wedge(\bvvar^{\cv}_\idx = k)
\end{align*}  

}
{
}
Applying this repair analysis to the examples in Figures~\ref{fig:client} 
and~\ref{fig:db}, \TarTar finds 13 admissible clock reference modification repairs, each involving two modifications.
Nine repairs exchange $y$ in the constraints $y \leq 1$ and $y \geq 1$ by a selection from the set of clocks
$z$, $x$ and $w$.
Four repairs exchange $y$ in the constraint $y \leq 1$ by $w$ or $x$, and $w$ in the constraint $w\geq 1$ by $y$ or $z$.

\paragraph{Reset Clock Repair Analysis.}
This analysis aims to repair a property violation by adding or removing clock resets.
We introduce a variation variable $\bvvar^{\rv}_{\idx,j}$ for
each clock $c_\idx$ and the transition leaving location $\lambda_j$ in the TDT.
The reset status in the extended constraint system is inverted when $\bvvar^{\rv}_{\idx,j} \neq 0$: 
if $c_i$ was not reset before, it will now be reset, and vice versa.
\ifthenelse{\boolean{arXiv}}
{
This is encoded by the clock reset variation constraints $\mathmbox{{\cal R}^{\rv}}$ and $\mathmbox{{\cal D}^{\rv}}$:
\begin{align*}  
\mathmbox{{\cal R}^{\rv}} \equiv \; &
\bigwedge
\limits_{c_i \in \mathmbox{\textit{reset}}(\lambda_j)}
  c_{i, j+1} = \left \{
  \begin{aligned}
    &0, && \text{if}\ \bvvar^{\rv}_\idx = 0 \\
    &c_{i} + \delta_j, && \text{otherwise}
  \end{aligned}\quad . \right .
\\
\mathmbox{{\cal D}^{\rv}} \equiv \; &
\bigwedge
\limits_{c_{i} \not\in \mathmbox{\textit{reset}}(\lambda_j)}
  c_{i, j+1} =  \left \{
  \begin{aligned}
   &c_{i} + \delta_j, && \text{if}\ \bvvar^{\rv}_\idx = 0 \\
    &0, && \text{otherwise}
  \end{aligned}\quad . \right.
\end{align*}  

}
{}
Applying the reset repair analysis to the examples in Figures~\ref{fig:client} 
and~\ref{fig:db}, \TarTar finds four admissible repairs. 
One repair removes the reset of clock $y$, another removes the reset of clock $z$ and two repairs add a 
reset of clock $x$ either on the transitions towards the state {\it reqProcessing} or the transition
towards the state {\it serReceiving}.

\paragraph{Urgent Location Repair Analysis.}
This analysis aims to repair cases where a
faulty usage of urgent locations, which are always left with zero delay after entering, causes a property violation.
Urgency of a location is modeled in the TDT constraint system by 
setting the location delay $\delta_j$ to $0$. We define a fresh variation variable 
$\bvvar^{\uv}_\idx$ for a location $\lambda_j$.
For $\bvvar^{\uv}_\idx \not= 0$, the urgency for a location $\lambda_j$ is inverted. 
\ifthenelse{\boolean{arXiv}}
{
We encode this idea using the following urgency variation constraint $\mathmbox{{\cal U}^{\uv}}$:
\begin{align*}  
\mathmbox{{\cal U}^{\uv}} \equiv \; &
\bigwedge\limits_{j \in \mathmbox{\text{\em urgent}(S)}}
 \bvvar^{\uv}_{i} = 0 \implies \delta_j = 0
\ \ \wedge
\bigwedge\limits_{j \not\in \mathmbox{\text{\em urgent}(S)}}
 \bvvar^{\uv}_{i} \not= 0 \implies \delta_j = 0.
\end{align*}  

}
{
}
Applying the urgency location repair analysis to the examples in Figures~\ref{fig:client} 
and~\ref{fig:db}, \TarTar finds two inadmissible repairs. 
The first one makes the state {\it reqAwaiting} urgent, and another repair makes 
the state {\it serReceiving} urgent.

\section{Usage of \TarTar}
We have implemented all repair analyses described in~\cite{KoeLeuWie19} and in this paper in a tool named \TarTar.
It provides a graphical user interface, a command-line interface and a web-interface 
which enables the execution of this resource intensive software on compute servers.
A user selects one of these interfaces via arguments provided when invoking the Java library 
implementing \TarTar. For real-time model checking, \TarTar relies on Uppaal.

\begin{figure}[htb]
\hfill
\subfigure[\TarTar GUI]{\includegraphics[width=0.4\textwidth]{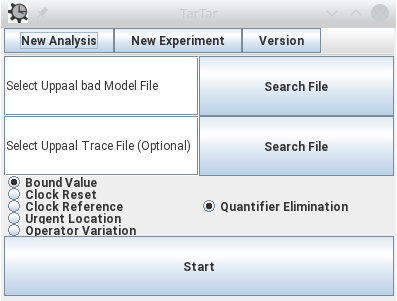}\label{fig:gui}}
\hfill
\subfigure[\TarTar Architecture]{\includegraphics[width=0.55\textwidth]{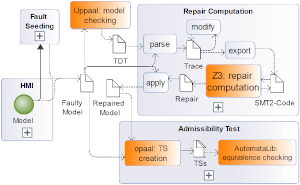}\label{fig:tartar}}
\hfill
\caption{\TarTar Tool}
\end{figure}

\begin{itemize}
    \item The argument {\it --web} launches the web server and corresponding interface. 
    \item Any other arguments launches the command-line mode. When using the argument {\it --help}, the command-line console  
    prints some help information.
    \item When no arguments are given, the graphical user interface depicted in Figure~\ref{fig:gui} is launched.
    The interface offers three tabs. {\it New Analysis} starts a repair analysis, {\it New Experiment} starts fault seeding that is described later in Section~\ref{sec:case_study}, and {\it Version} shows the current version number of \TarTar.
\end{itemize}

All tool interfaces expect identical inputs in order to start a \TarTar analysis run.
The user specifies a file containing the Uppaal model as input and selects the kind of repair to compute. Optionally, a file with a TDT of the given Uppaal model can be specified. When no TDT is provided, \TarTar automatically calls Uppaal to compute a TDT.
The result of an analysis is one repaired model file for every computed repair, as well as a text file that summarizes which repairs are admissible.

\section{Software Architecture and Implementation of \TarTar}

The software architecture of \TarTar is depicted in Figure~\ref{fig:tartar}. The orange rectangles in the figure represent external tools that \TarTar calls in the course of the repair analysis.
Uppaal is a state-of-the-art and closed-source model checking tool, which \TarTar uses to compute a TDT for a given model and property.
The SMT solver Z3~\cite{deMBjo08} is used to solve the generated partial MaxSMT problems. To check the admissibility of a repair, \TarTar uses 
opaal and the AutomataLib component of LearnLib~\cite{IsbHowSte2015} since they conveniently provide
functionality used during admissibility checking.

\paragraph{Data Flow Architecture.}
\TarTar consists of many computation steps. For example, a TDT is parsed internally and stored as a Trace. This Trace is then modified and exported as SMT-LIB2~\cite{smtlib17} code.
We define a computation step of \TarTar as the computation transforming input into result artifacts.
This focus on artifacts
ensures a highly cohesive architecture and clear interfaces between any two computation steps.
Computation steps with identical objectives are grouped into a project. 
This results in four projects depicted by blue rectangles in Figure~\ref{fig:tartar}.
\begin{itemize}
    \item {\it HMI} denotes the user interfaces of \TarTar. The user inputs a timed model. \TarTar then calls the projects {\it Repair Computation} 
    using a faulty timed model as a parameter.
    In case that the model is correct, \TarTar calls the project {\it Fault Seeding}.
    \item {\it Fault Seeding} seeds faults into a correct model and then repairs the faulty model by computing repairs using {\it Repair Computation}. 
    We use this analysis in Section~\ref{sec:case_study} in order to benchmark the {\it Repair Computation} analyses.
    \item {\it Repair Computation} computes candidate repairs for a faulty timed model, applies these repairs
    to the model
    and finally automatically calls the {\it Admissibility Test}.
    \item {\it Admissibility Test} checks for every repaired model whether the computed repair is also admissible. 
\end{itemize}

\paragraph{Control Flow Architecture.}
\TarTar computes iteratively a set of repairs for a given faulty Uppaal model and a given property $\Pi$ 
using the following steps:

\begin{enumerate}
\item[0.] {\em Counterexample Creation}. \TarTar calls Uppaal
to verify the model against $\Pi$. In case $\Pi$ is violated, it stores a shortest symbolic
TDT witnessing the violation in XML format. 
\item {\em Diagnostic Trace Creation}. 
\TarTar parses the model and the TDT into a data structure {\it Trace}.
To add potential repairs, \TarTar copies the trace and replaces the constraints that will 
potentially be subject to a repair by their modified variants. 
The modified trace is then translated to a logic constraint system, represented in SMT-LIB2 code.

\item {\em Repair Computation}. 
Z3~\cite{deMBjo08} then solves a MaxSMT problem on the modified trace constraint system, 
computing a repair in which the
number of unmodified constraints on the variation variables of type $\bvvar=0$ is maximized.
Since Z3 can solve a MaxSMT problem only for quantifier-free linear real arithmetic, 
\TarTar first runs a quantifier elimination on the constraint system. It then 
solves the MaxSMT problem with soft constraints requiring $\bvvar=0$ for all variation variables.
\ifthenelse{\boolean{arXiv}}
{}
{
For a more comprehensive presentation of this construction we refer the reader to~\cite{KoeLeuWie20}.
}
In case no solution is found, \TarTar terminates.
Otherwise, \TarTar applies the repair to the faulty model and returns a repaired model.

\item {\em Admissibility Check}. \TarTar checks the admissibility of a repair and compares the untimed languages of the faulty and repaired models. 
\TarTar calls the model checker opaal in order to compute the timed transition systems (TTS) of the original and the repaired Uppaal model.
We modified the opaal model checker in such a way that it returns the TTS for a model.
\TarTar then checks whether the two TTS have equivalent untimed languages, in which case the repair is admissible. 
This check is implemented using the library AutomataLib.
In case the two TTS are not equivalent, the admissibility test returns
a trace as a witness for the difference.

\item {\em Iteration}.
\TarTar enumerates all repairs, i.e., all combinations of constraint modifications that correct the TDT.
The repairs are iteratively enumerated starting with the ones that require the smallest number of modifications to the model.
After a repair is computed, 
the combination of modified variables that has been found 
is prevented from being reconsidered for future repairs by setting these modification variables to 
$0$ using hard asserts.
\TarTar then proceeds with attempting to compute further, previously unconsidered repairs.
\end{enumerate}

\paragraph{Component Architecture.}
\begin{wrapfigure}{R}{0.50\textwidth}
\centerline{
\includegraphics[width=0.50\textwidth]{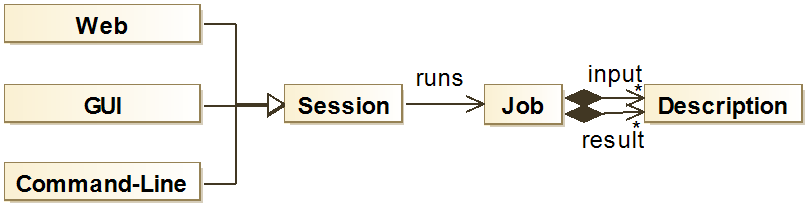}
}
\caption{\TarTar Component Architecture}
\label{fig:infra}
\end{wrapfigure}

We implemented \TarTar with the general infrastructure depicted in Figure~\ref{fig:infra}.
The interface {\it Job} provides a general abstraction for an algorithm and specifies the necessary input and result 
values of the algorithm by the class {\it Description}.
\TarTar contains a {\it Job} for the projects {\it Fault Seeding}, {\it Repair Computations} and {\it Admissibility Test}.
The class {\it Session} executes a {\it Job} and derivations of {\it Session} provide the different interfaces to the user.
With this infrastructure, the analysis implementation in \TarTar is independent from the implementation 
of the user interfaces, thus reducing coupling and improving modifiability of the code.

\paragraph{Implementation Details.}
We implemented the different projects that constitute \TarTar in Java and use 
the build-management tool maven~\cite{mvn} to manage the dependencies between the projects.
\TarTar interacts differently with the external tools that are needed for different purposes.
It calls Uppaal via the command-line interface in order to
generate a TDT, calls Z3 via its API to compute a repair. For the admissibility check, it calls opaal using a command-line script and the AutomataLib as an included Java library.
For the implementation of the \TarTar analyses the following two details are essential.

We modify constraints in an Uppaal model in order to apply a repair or to seed a fault. 
Since neither clock constraints nor transitions possess explicit unique identifiers in an Uppaal model, 
it is not obvious which constraint to change.
We therefore uniquely identify a constraint by traversing the constraints in the sequence stored in the 
Uppaal model file and use the constraint index in this sequence as its identifier.

The complexity of the algorithms for solving quantifier elimination and the MaxSMT problem increase 
exponentially with the number of variables in the SMT model~\cite{KoeLeuWie19}. 
We therefore reduce the number of variables by exploiting implied equality constraints. For example, a variable $c_j$ is created for every clock $c$ in every step $j$ of the TDT. We eliminate $c_j$ explicitly before quantifier elimination by replacing it with the term $\sum\nolimits_{i \in r..j}d_i$, where $d_i$ is the time delay at step $i$ in the trace and $r$ is the last step before $j$ where $c$ was reset.

 \section{Evaluation}\label{sec:case_study}

\paragraph{Evaluation Strategy.}
In order to evaluate the repair analyses both qualitatively and quantitatively, 
we need to synthesize a set of faulty timed automata.
To the best of our knowledge, no benchmark suite for faulty timed automata exists.
We therefore create faulty models by using the fault seeding strategy from~\cite{KoeLeuWie19} 
which is motivated by ideas from mutation testing~\cite{JiaHar2011}.
Mutation testing evaluates the quality of a test suite for a given program by systematically corrupting program code and determining the ratio of corruptions that the test suite is able to detect. We apply the same principle to evaluate the quality of our repair technique. 
As proposed in~\cite{KoeLeuWie19}, 
fault seeding modifies a single clock constraint so that the result is a set of models that violate a given property.
During the seeding, the bound of a single clock constraint is modified by an amount of $\{-10, -1, +1, +0.1M, +M\}$, 
where M is the maximal clock bound occurring in a given model.
Our observation was that making either small modifications that are close to the bound value or
modifications in the order of the maximal bound value M often introduce actual errors in the model.
We have extended fault seeding to the new types of repairs. 
In particular, fault seeding additionally exchanges the comparison operator in a clock constraint by $\{<, \leq, =, \geq, >\}$, 
swap a referenced clock with all other clocks occurring in the given model, 
modify the reset clocks of any transition, 
and switch for any location whether it is urgent.
\TarTar checks automatically whether a modified TA violates a given property.
If this is the case, it performs all of the above defined repair analyses.

\paragraph{Results.}
We applied fault seeding to the models in~\cite{KoeLeuWie19} and analyzed the obtained TDTs using 
the above described repair analyses implemented in \TarTar. All analyses were performed on a computer with an i7-6700K CPU (4.00GHz), 60GB of RAM and a 64 bit Linux operating system. We summarize the results of the experiment per considered model (Table~\ref{tab:result_model}) and per type of considered repair (Table~\ref{tab:result_repair}).
\begin{table}[b]
{\centering
\begin{scriptsize}
\begin{tabular}{|c||r|r|r|r|r|r|r|r|r|r|r|r|r|r|r|}
\hline
\multicolumn{1}{|c||}{Repair} &	\multicolumn{1}{c|}{\#Sd}	&		
\multicolumn{1}{c|}{\#T} & \multicolumn{1}{c|}{$T_{\textit{UP}}$}&
\multicolumn{1}{c|}{\textit{Ln}}	&	\multicolumn{1}{c|}{\textit{\#R}}	&
\multicolumn{1}{c|}{\textit{\#A}}	&	\multicolumn{1}{c|}{\textit{\#S}}	&
\multicolumn{1}{c|}{$T_{\textit{QE}}$}	&	\multicolumn{1}{c|}{\#O} & \multicolumn{1}{c|}{$T_{\textit{R}}$}	&
		
\multicolumn{1}{c|}{$M_R$}	& \multicolumn{1}{c|}{\textit{\#Vr}}	&	\multicolumn{1}{c|}{\textit{\#Cn}} &	\multicolumn{1}{c|}{$T_{\textit{Adm}}$ }
 &	\multicolumn{1}{c|}{$M_A$ }
\\
\hline \hline
db rep. & 110 & 13 & 0.016 & 4 & 229 & 138 & 9 & 89.346 & 2 & 0.911 & 14.53 & 30 & 91 & 2.080 & 45\\
csma & 191 & 10 & 0.012 & 2 & 70 & 26 & 8 & 0.049 & 0 & 0.023 & 0.58 & 16 & 72 & 1.825 &  75\\
elevator &88 & 5 & 0.011 & 1 & 7 & 5 & 4 & 0.049 & 0 & 0.020 & 0.53 & 6 & 28 & 1.665 & 17\\
viking & 310 & 9 & 0.015 & 18 & 9 & 7 & 5 & 86.539 & 21 & 1.436 & 20.07 & 120 & 180 & 1.952 & 543\\
bando & 1,955 & 40 & 0.111 & 279 & 4,061 & 209 & 21 & 31.555 & 46 & 4.922 & 20.86 & 1,156 & 8,144 & 19.57 & 1251\\
Pacemaker & 1,187 & 12 & 0.022 & 9 & 62 & 19 & 10 & 0.663 & 20 & 0.325 & 2.59 & 116 & 988 & 1.994 & 206\\
SBR & 353 & 50 & 0.027 & 84 & 751 & 660 & 31 & 117.057 & 86 & 2.686 & 37.16 & 765 & 1,211 & 138.004 & 211\\
FDDI & 314 & 36 & 0.014 & 11 & 166 & 105 & 34 & 29.859 & 51 & 3.074 & 9.70 & 116 & 272 & 2.241 & 128\\ 
\hline
\end{tabular}
\end{scriptsize}
\vspace{0.2cm}
\caption{\label{tab:result_model} Experimental results according to model.} }
\end{table}
Column {\it Sd} contains the count of seeded faults that result in a number {\it \#T} of faulty models. {\it $T_{\textit{UP}}$} is the maximal time that Uppaal needs to create a TDT for the faulty models, and the longest TDT has a length of {\it Ln}. \TarTar computed for the TDTs overall a number {\#\it R} repairs of which {\#\it A} are admissible. An admissible repair is found for {\#\it S} of the TDTs.
The computation effort for a repair analysis is given by the time {\it $T_{\textit{QE}}$} for successful quantifier elimination, the number of timeouts {\#\it O} of quantifier eliminations after 10 minutes, the average time {\it $T_{\textit{R}}$} to compute a repair and the memory consumption {\it $M_R$}.
The constraint system that Z3 solves has the count {\#\it Vr} of variables and {\#\it Cn} of constraints.
The effort for the admissibility check is given in time {\it $T_{\textit{Adm}}$} and memory {\it $M_A$}. 
All times are given in seconds and memory consumption in MB.
Notice that we omit the columns pertaining to the fault seeding and TDT computation in Table~\ref{tab:result_repair} as they are irrelevant here.

Overall, \TarTar seeded 4.508 faults. This resulted in 175 TDTs in total ($60$ TDTs due to bound modification, 
$72$ due to operator variation, $27$ due to changing the clock reference, $8$ due to complementing the reset of clocks 
and $8$ due to the switching of urgent locations).
\TarTar found 5,355 repairs, out of which 1,169 were admissible. 
It found at least one admissible repair for 122 of the TDTs. The maximal number of modified constraints in the admissible repairs computed for a single TDT using all types of analysis was $25$.

\begin{table}[tb]
{\centering
\begin{scriptsize}
\begin{tabular}{|c||r|r|r|r|r|r|r|r|r|r|r|r|}
\hline
\multicolumn{1}{|c||}{Repair} &	\multicolumn{1}{c|}{\textit{\#R}}	&
\multicolumn{1}{c|}{\textit{\#A}}	&	\multicolumn{1}{c|}{\textit{\#S}}	&
\multicolumn{1}{c|}{$T_{\textit{QE}}$} &	\multicolumn{1}{c|}{\#O}	&	\multicolumn{1}{c|}{$T_{\textit{R}}$}	&
		
\multicolumn{1}{c|}{$M_R$}	& \multicolumn{1}{c|}{\textit{\#Vr}}	&	\multicolumn{1}{c|}{\textit{\#Cn}} &	\multicolumn{1}{c|}{$T_{\textit{Adm}}$ }
 &	\multicolumn{1}{c|}{$M_A$ }\\
\hline \hline
Bound Modification & 533 & 364 & 85 & 15.209 & 8 & 4.922 & 20.86 & 1,156 & 2,498 & 138.004 & 525\\
Operator Variation & 3,929 & 96 & 51 & 117.057 & 44 & 2.686 & 37.16 & 996 & 8,144 & 59.117 & 543\\
Clock Reference & 693 & 625 & 35 & 33.282 & 61 & 3.074 & 14.13 & 1,120 & 5,355 & 116.944 & 206\\
Reset Clock & 45 & 37 & 13 & 89.346 & 113 & 0.911 & 14.53 & 996 & 2,836 & 2.051 & 45\\
Urgent Location & 155 & 47 & 37 & 0.107 & 0 & 0.135 & 3.16 & 1,120 & 2,502 & 58.551 & 1,251\\
\hline
\end{tabular}
\end{scriptsize}
\vspace{0.2cm}
\caption{\label{tab:result_repair} Experimental results according to type of repair.} }
\end{table}

\paragraph{Interpretation.}
Few of the seeded faults resulted in a property violation. \TarTar seeded 4.508 faults which led to $175$ TDTs, thus only 3.9\% of these faults result in a TDT. This supports the hypothesis that, in practice, often times only few time constraints have an impact on a property violation.
\TarTar computes at least one admissible repair by bound modification for $85$ (48\%) of the $175$ TDTs, by operator variation for $51$ (29\%), by clock reference for $35$ (20\%), by clock reset for $13$ (7\%) and by urgent location for $37$ (21\%).
Every analysis on its own computes less admissible repairs than the combination of all repair analyses, which solves 
$122$ (69\%) of the $175$ TDTs.
The largest number of modified constraints in all the admissible repairs for a single TDT was $25$, which is less than anticipated. 
This low number of modified constraints infer that, for the examples that we considered, only a few constraints of each TDT combined to admissible repairs.
The number of modified constraints determines the number of possible repairs that have an impact on whether a property is violated or not.
Since it was observed in~\cite{KoeLeuWie19} that the computational effort for the repair computation is largely 
determined by the quantifier elimination step, we expect that in light of the observed $226$ timeouts 
a more efficient quantifier elimination would lead to a significantly higher number of repairs.
Furthermore, the number of timeouts, and thus the computation time needed for the repair, rises with the length of the analyzed TDT.
The model {\it SBR}  has the most timeouts with $86$ and the third longest trace with a length of $84$ steps. 
The model {\it bando} has the third most timeouts with $46$ and the longest trace.
Obviously, the longer the TDT, the larger the resulting constraint system, leading to increased computational effort.
The {\it bando} model has the largest constraint system with $1,156$ variables and $8,144$ constraints. 
The {\it SBR} model has the second largest constraint system with $765$ variables and $1,211$ constraints.
The model {\it FDDI} has a shorter trace of length of $11$ and a much smaller constraint system with $116$ variables 
and $272$ constraints.
From this we conclude that the complexity of a repair depends not only on the trace length, but also on the 
intrinsic complexity of the model.
Modifying states from urgent to non-urgent during fault seeding resulted in only $8$ TDTs. 
This low number is due to the observation that the considered models contain only few urgent states. 
Modifying non-urgent states to urgent ones, however, did not lead to a single property violation
resulting in a TDT.
The rationale is that urgency ensures to leave a state immediately without a delay which leads to a restriction 
rather than a relaxation regarding the time budget spent along an execution trace.
As a consequence, making a state urgent does not cause a property violation in many models since
the type of the checked properties is typically time bounded reachability, and a restricted 
time budget does not make it more likely that the property is violated.
We finally observe that the admissibility check requires more computation resources than the repair computation. 
The maximal memory used for the admissibility test was $1,251\text{MB}$ in contrast to $37.16\text{MB}$ for the repair computation.
This is in line with our expectation since the admissibility test searches the state space of the full NTA, 
while the repair analyses only considers a single TDT.
 \section{Conclusion}\label{sec:conclusion}
We have presented the \TarTar tool, its architecture and implementation, 
and illustrated its application to a number of significant case studies.
In the course of our work we have extended the repair analysis that is implemented in \TarTar for bound modification to modifications of comparison operators, clock references, reset of clocks and missing urgencies.
The evaluation of the repair analyses showed that an admissible repair is computed for at least 69\% of the analyzed TDTs.
The integration of various tools with heterogeneous interfaces posed a particular challenge to the architecture
of \TarTar which we addressed by the definition of intermediate artifacts.

 In future work we plan to explore the interplay between different repairs that are computed for a repaired system that still violates a property, and develop refined strategies to select promising repairs from a repair set.
 A further generalization of the analysis is to not only compute clock constraint modifications for faulty models but also 
to compute possible relaxations of clock constraints for correct models in order to support design space exploration.

\bibliographystyle{alpha}
\bibliography{bibliography}

\end{document}